\documentclass[twocolumn,showpacs,secnumarabic,amssymb,nobibnotes,aps,prl]{revtex4-1}
\usepackage{amssymb}
\usepackage{graphicx}
\begin{document}
\title[Initial Correlation Dependence of Aging]{Initial Correlation Dependence of Aging in Phase Separating Solid Binary Mixtures and Ordering Ferromagnets}
\author{Subir K. Das$^1$}
\email{das@jncasr.ac.in}
\author{Koyel Das$^1$}
\author{ Nalina Vadakkayil$^1$}
\author{ Saikat Chakraborty$^{1,\,2}$}
\author{Subhajit Paul$^{1,\,3}$}
\affiliation{$^1$Theoretical Sciences Unit and School of Advanced Materials, Jawaharlal Nehru Centre for Advanced Scientific Research,
Jakkur P.O., Bangalore 560064, India.}
\affiliation{$^2$Institut f\"{u}r Physik, Johannes-Gutenberg Universit\"{a}t Mainz, Staudinger Weg 9, 55099 Mainz, Germany}
\affiliation{$^3$Institut f\"{u}r Theoretische Physik, Universit\"{a}t Leipzig, Postfach 100920, D-04009 Leipzig, Germany}
\begin{abstract}
 Following quenches of initial configurations having long range spatial correlations, prepared at the demixing 
 critical point, to points inside the miscibility gap, we study aging phenomena in solid binary mixtures. Results on the decay of the two-time order-parameter  autocorrelation functions, obtained 
from Monte Carlo simulations of the two-dimensional Ising model, with Kawasaki exchange kinetics, are analyzed via state-of-the art methods. The outcome is compared with 
 that obtained for the ordering in uniaxial ferromagnets. For the latter, we have performed Monte Carlo simulations of the same model using the Glauber mechanism. 
 For both types of systems we provide comparative discussion of our results with reference to those concerning quenches with configurations having no spatial correlation. We also discuss the role of structure on the decay of these correlations.
\end{abstract}
\maketitle
\section{Introduction}
Having been prepared at a high starting temperature ($T_s$), when a homogeneous mixture is quenched to a final temperature ($T_f$), that falls inside the miscibility gap, 
it renders unstable to fluctuation and separates into regions or domains rich in particles of similar type \cite{binder,bray,onuki,ral,gunton}. Kinetics of such phase separation is of immense interest from both scientific and technological viewpoints. To probe the aging during such evolution, often one studies the decay of the  two-time auto-correlation function \cite{mz}
\begin{eqnarray}\label{auto_corr_fn}
 C_{\textrm{ag}}(t,t_w)=&\langle\psi(\vec{r},t)\psi(\vec{r},t_w)\rangle-\nonumber\\
 &\langle{\psi(\vec{r},t)}\rangle \langle\psi(\vec{r},t_w)\rangle.
\end{eqnarray}
Here $\psi$, chosen scalar by keeping the content of the paper in mind, is a space ($\vec{r}$) and time dependent order parameter, while $t$ and $t_w$ ($\leq t$) are referred, respectively, 
to as the observation and waiting times.
\par
Due to the violation of time-translation invariance in nonequilibrium systems, $C_{\textrm{ag}} (t,t_w)$ for different $t_w$ are not equivalent to each other. In other words, if this correlation function is plotted versus $t-t_w$, there will be no collapse of data for different values of $t_w$. However, it is found that in many systems $C_{\textrm{ag}}(t,t_w)$ exhibits the scaling behavior \cite{mz,fisher_huse,liu,yrd,henkel,corberi,lorenz,midya_jpcm,midya_pre,nv,paul,bera,corberi_villa,humayun,bray_humayun,dutta,puri_kumar}
\begin{equation}\label{scaling_auto_corr}
    C_{\textrm{ag}}(t,t_w) \sim (\ell/\ell_w)^{-\lambda},
\end{equation}
where $\ell$ and $\ell_w$ are the average sizes of domains at times $t$ and $t_w$, respectively. Note that $\ell$ typically has a power-law time dependence \cite{binder,bray,onuki,ral,mz}
\begin{equation}
    \ell \sim t^\alpha,
\end{equation}
in phase ordering systems. Here $\lambda$ and $\alpha$ are referred to as the aging and growth exponents. Values of these exponents, along with few other properties \cite{bray,bray_majumdar}, define the 
nonequilibrium universality classes \cite{bray,humayun}.
\par
It has been argued that for same model, depending upon the spatial correlation in the initial configurations there can be different universality classes \cite{humayun,bray_humayun,dutta} -- one for $T_s=\infty$ 
and the other for $T_s=T_c$, the latter being the critical temperature. Note here that at $T_{s}=\infty$ a system, in standard picture, has a correlation length $\xi=0$ and at $T_s=T_c$, $\xi=\infty$, 
when the system is of thermodynamically large size \cite{bray,onuki,fisher}. For ordering in uniaxial ferromagnets \cite{bray,fisher}, this fact of universality has been demonstrated \cite{humayun,bray_humayun}. There the understanding is that even though $\alpha$ remains the same \cite{humayun}, $\lambda$ and other 
dynamic and structural quantities are different in the two classes \cite{humayun,bray_humayun,saikat_epjb,saikat_pre,blanchard}.
\par
In contrast to the magnetic case, for which there is no constraint on the conservation of system integrated order parameter during evolution \cite{bray}, the task of understanding of coarsening phenomena is known to be significantly more difficult, at least theoretically and computationally, for conserved order parameter dynamics that 
applies to kinetics of phase separation  in multi-component mixtures \cite{bray}. Computational difficulty \cite{amar,sm_2010,sm_2011}, to a certain extent, arises from the significantly slower dynamics in the latter case. Note that for the nonconserved case \cite{bray,allen_cahn} $\alpha = 1/2$, whereas for the conserved case \cite{bray,lifshitz,huse} $\alpha=1/3$. Furthermore, irrespective of the type of dynamics, conserved or nonconserved, quantitative understanding of aging behavior, even for simple models, still remains difficult, convergence in the settlement of issues being rather slow \cite{fisher_huse,liu,yrd,henkel,corberi,lorenz,midya_jpcm,midya_pre,nv,paul,bera,corberi_villa,humayun,bray_humayun,bray_majumdar,marko,sm_2013,ahmad,yeung_jasnow}, despite the availability of huge computational resources.
\par
Nevertheless, significant progress has recently been made, following adoption of methods of analysis that are analogous to the popular techniques used for extracting information about equilibrium systems. In a recent work \cite{midya_pre} we have quantified the values of $\lambda$ for phase separating binary mixtures (A+B) in different space dimensions $d$, via formulation and application of finite-size scaling technique \cite{midya_jpcm,fisher_barber} 
to Monte Carlo (MC) simulation results, for quenches with initial $\xi=0$. For this and a number of other situations, including the ferromagnetic case, we have demonstrated \cite{midya_jpcm,midya_pre,nv,paul,bera} that $\lambda$ satisfies certain bounds. 
Here note that Fisher and Huse (FH) argued \cite{fisher_huse}:
\begin{equation}\label{fh_bound}
    \lambda \geq \frac{d}{2}.
\end{equation}
 Later, Yeung, Rao and Desai (YRD) \cite{yrd} provided a more accurate and generic bound:
\begin{equation}\label{yrd_bound}
\lambda \geq \frac{d+\beta}{2},
\end{equation}
where $\beta$ is an exponent related to the short wave number ($k$) behaviour of structure factor \cite{yeung}, viz.,
\begin{equation}\label{powerlaw_beta}
 S(k \rightarrow 0, t_w) \sim k^\beta.
\end{equation}
For random initial configurations ($\xi=0$), $\beta=0$ and so, the YRD bound coincides with that of FH. For nonconserved order parameter, when $T_s=\infty$, $\beta=0$ even in the long time limit. The latter, however, 
is not true for the conserved case \cite{yeung,furukawa,majumdar_huse}. This is one of the reasons for our observation of vastly different $\lambda$ values in the two cases, irrespective of space dimension, for quenches with $\xi=0$.

When started from $T_s=T_c$, it is expected that one will have different structural scaling \cite{humayun, bray_humayun}. If so, the bounds for both conserved and nonconserved order-parameter will be different 
from that when quenched from $T_s = \infty$. This provides an intuitive understanding that $\lambda$ for both conserved and nonconserved classes will be different for $T_s = \infty$ and $T_s = T_c$, giving rise to different universalities. 
This is demonstrated, as already stated, theoretically and computationally, for the non-conserved case \cite{humayun,bray_humayun}.

In this paper we focus on the conserved case, i.e., we take up the task of estimating $\lambda$ for binary (A+B) mixtures with $T_s = T_c$. Note that nonequilibrium universality classes are also decided \cite{bray} by the space dimension, 
order-parameter symmetry and presence of hydrodynamics. In this paper we focus on $d=2$  and scalar order parameter, in absence of hydrodynamics, i.e., in our model system coarsening occurs due to simple diffusive transport, 
as expected in solid alloys. To validate our method, and thus, the result, we also estimate $\lambda$ for the nonconserved case that can be readily compared with the existing results from other approaches \cite{humayun,bray_humayun}. 

We show that the obtained values of $\lambda$ are consistent with the YRD bound. These numbers are discussed with reference to the corresponding numbers \cite{midya_pre} for $T_s=\infty$. It transpires that for conserved order parameter also $\lambda$ for 
$T_s = T_c$ is hugely different from that for $T_s=\infty$. Another recent study of ours \cite{nv_2} suggest that in both the cases the growth exponent remains same, like in the nonconserved case. 
Thus, there exists qualitative similarity between conserved and nonconserved cases, with respect to relaxation following quenches to the ordered region.
\section{Models and Methods}
We study nonequilibrium dynamics in solid binary mixtures and uniaxial ferromagnets, via Kawasaki exchange \cite{kawasaki} and Glauber spin flip \cite{glauber} Monte Carlo methods \cite{binder_heermann,landau,dan_frenkel}, respectively, using the Ising model \cite{fisher} on a 2D square lattice, with periodic 
boundary conditions \cite{landau} in both the directions. 
The Hamiltonian of the model is given by \cite{fisher,landau} 
\begin{equation}\label{hamilton}
 H=-J\sum\limits_{<ij>} S_i S_j; \, S_i=\pm 1; \,J>0,
\end{equation}
where the values $+1$ and $-1$ correspond to particles of types A and B, respectively, in the case of binary mixture, and up and down spins in the case of ferromagnet. The value of critical 
temperature of this model \cite{fisher,landau} in $d=2$ is $\simeq 2.269J/k_B$, where $J$ is the interaction strength and $k_B$ is the Boltzmann constant.

A trial move in the Kawasaki exchange Ising model (KIM) is the interchange of particles between randomly 
selected nearest neighbor sites, whereas in the Glauber Ising model (GIM) a move is performed by flipping a randomly selected spin. In both the cases the probability of acceptance of 
trial moves is given by \cite{binder_heermann,landau,dan_frenkel}
\begin{equation}\label{metropolis}
 P(i \rightarrow j) = \textrm{min} (1,\exp(-(E_i-E_j)/k_B T_f)),
\end{equation}
where $E_{i(j)}$ is the energy of the state $i(j)$. Time in our simulation is estimated in units of MC
steps (MCS), where one MCS is equivalent to $L^2$ trial moves, $L$ being the linear dimension of a square box, in units of the lattice constant $a$. In the rest of the paper, we set $J$, $k_B$ and $a$ to unity. 

Unless otherwise mentioned, we quench the systems from $T_s = T_c^L$, $T_c^L$ being the system-size dependent critical temperature \cite{fisher_barber,luijten}, to a final temperature $T_f=0.6T_c$. 
In order to obtain the equilibrium configurations at $T_c^L$, we have performed simulations using Wolff algorithm \cite{wolff}, that, to a good degree, helps avoiding the critical slowing down \cite{hohenberg}. 
Here, instead of a single spin, a randomly selected cluster of similar particles or spins is flipped.

The average domain lengths of a system 
during evolution have been calculated via \cite{sm_2010,sm_2011}
\begin{equation}
 \ell(t)=\int P(\ell_d,t)\ell_d d\ell_d,
\end{equation}
where $P(\ell_d,t)$ is a domain-size distribution function, and $\ell_d$ is the distance between two successive interfaces in a specific direction. 
In the calculation of the autocorrelation functions [see Eq. (\ref{auto_corr_fn})], the order parameter $\psi$ at a space point corresponds to the value of spin in Eq. (\ref{hamilton}) at a lattice site. All the presented results, for both KIM and GIM, are averaged over a minimum of $100$ independent initial configurations. 
\section{Results}
\begin{figure}
\centering
\includegraphics*[width=0.4\textwidth]{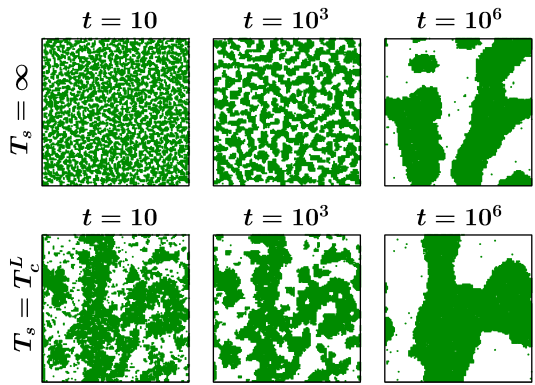}
\caption{\label{fig1} 
Snapshots for the Kawasaki Ising model during evolutions following quenches from $T_s=\infty$ (upper frames) and $T_s = T_c^L$ (lower frames), with $L=128$. In each of the cases pictures from three different 
times are shown. The dots represent A particles and the rest of the space is occupied by B particles. Here and in other places all results are from quenches to $T_f = 0.6T_c$.}
\end{figure}
We start by presenting results from the KIM. In Fig. \ref{fig1} snapshots during the evolutions for different $T_s$ values are presented. The upper frames are for $T_s = \infty$ and the lower ones are for $T_s = T_c^L$. All the pictures are for $L=128$. The difference in structure in the two cases is recognizable, even though there exist strong finite-size effects in the initial configurations  \cite{landau,fisher_barber} for $T_s=T_c^L$. The latter is in addition to the standard finite-size effects \cite{sm_2010,sm_2011,heermann} that is observed for $T_s=\infty$, 
when $\ell$ approaches $L$. As is well known \cite{fisher},
\begin{equation}\label{powerlaw_xi}
 \xi \sim \epsilon^{-\nu};\, \epsilon = \frac{T_s-T_c}{T_c},
\end{equation}
$\nu$ being a static critical exponent. For a true phase transition, achievable in thermodynamically large systems, of course, $\xi =\infty$ at the critical point. However for $L<\infty$, which is always the case for computer simulations, $\xi$ is finite, the maximum attainable value being $\xi=L$. 
Because of that, for finite $L$, when $T_s=T_c^L$, following quenches the systems quickly deviate from the desired \cite{humayun,bray_humayun} scaling form, different from that for quenches with $T_s=\infty$, of the nonequilibrium structure. This can be realized by taking a closer look at the snapshots for $T_s = T_c^L$ in Fig. \ref{fig1} -- the fractality is changing with time. This additional finite-size effect must be taken care of via appropriate extrapolation of the size-affected  
quantitative data in the $L=\infty$ limit. This requires knowledge of $T_c^L$ for various values of $L$. Related results we present next before showing data for the autocorrelation functions.
\begin{figure}
\centering
\includegraphics*[width=0.4\textwidth]{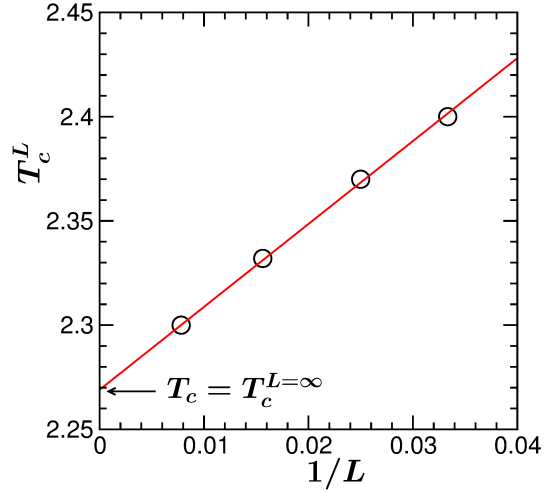}
\caption{\label{fig2} 
Plot of finite-size critical temperatures $T_c^L$ as a function of the inverse system size $1/L$. These results were obtained for GIM. The continuous line is a fit of the data set to the scaling form in Eq. (\ref{powerlaw_tcl}), by fixing $T_c$ and $\nu$ to their 2D Ising values. Unless otherwise mentioned, all the results below will correspond to $T_s=T_c^L$.}
\end{figure}
\par
Phase behavior for a model can be obtained via computer simulations by calculating the temperature dependent, appropriately defined, order-parameter distribution functions \cite{landau,luijten}. Such a phase diagram or coexistence curve will always suffer from finite-size effects due to the fact that, as mentioned above, in simulations we always have $L < \infty$. Nevertheless, via the applications of well-established scaling principles phase behavior, including the critical point, in the thermodynamic limit, can be satisfactorily obtained \cite{luijten,roy_das,midya_jchem}.  

In the two-phase or coexistence region the order-parameter distribution will have a double peak structure, locations of the peaks representing points along the coexistence curve. On the other hand, in the homogeneous (one-phase) region these distributions will have single peak shape (with temperature dependent width). The temperature 
at which the crossover from double peak to single peak structure occurs is identified as the value of $T_c^L$. 

A plot of $T_c^L$ versus $1/L$ is shown in Fig. \ref{fig2}. These results were obtained from GIM. Given that static critical universality is very robust, we will use the same data for the study of nonequilibrium phenomena in KIM as well. Note that the results for $T_c^L$ are expected to satisfy 
the scaling form \cite{landau,luijten,roy_das,midya_jchem}
\begin{equation}\label{powerlaw_tcl}
 T_c^L - T_c \sim L^{-1/\nu},
\end{equation}
validity of which can be checked from its consistency with Eq. (\ref{powerlaw_xi}). For the Ising model (universality class) $\nu = 1$ in $d=2$. The data set in Fig. \ref{fig2}, thus, is in agreement with this expected critical point behavior. Note that the continuous line in Fig. \ref{fig2} is a fit of the simulation data set to the scaling form in Eq. (\ref{powerlaw_tcl}), by fixing $\nu$ and $T_c$ to the 2D Ising values.
\begin{figure}
\centering
\includegraphics*[width=0.4\textwidth]{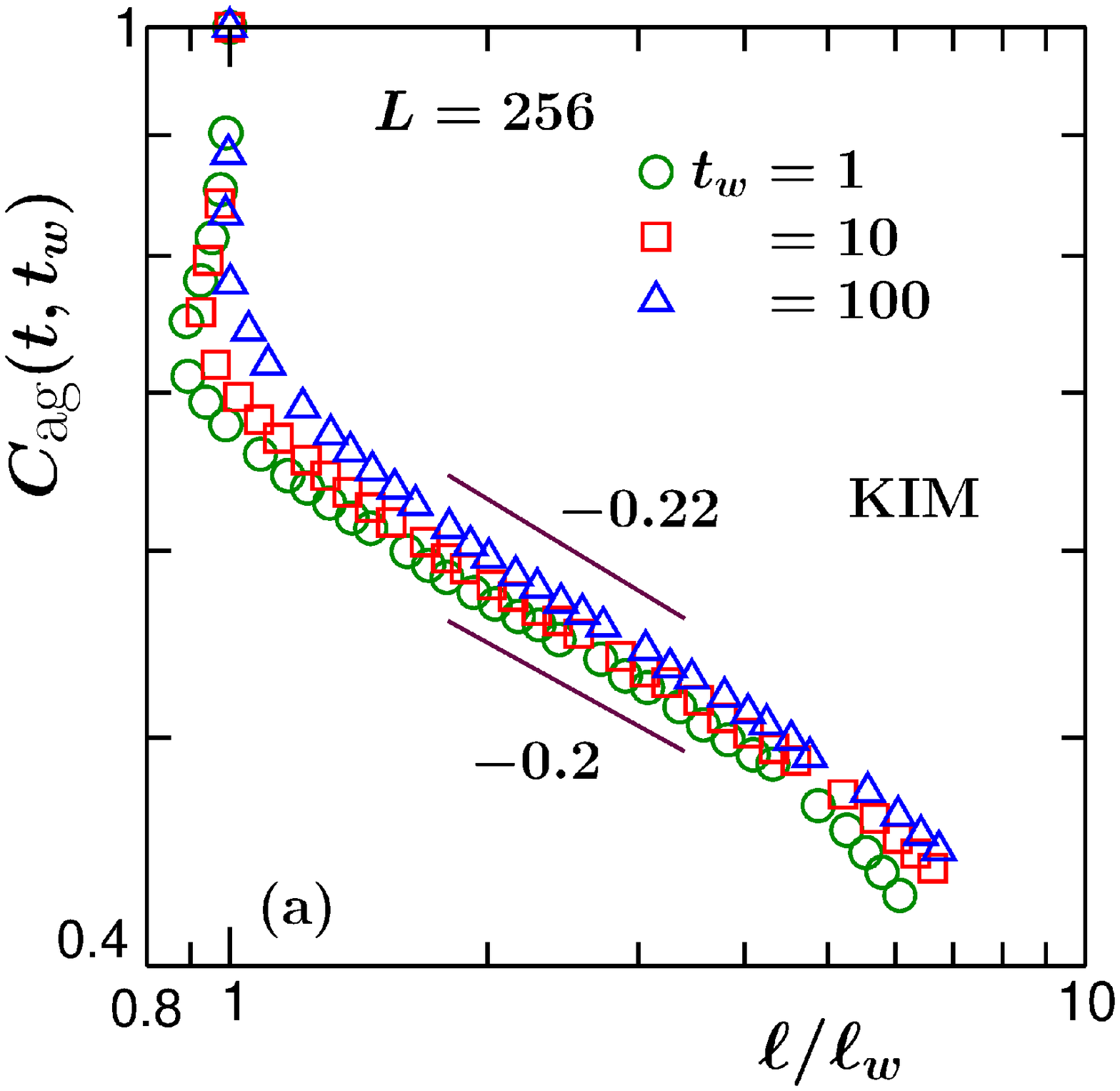}
\vskip 0.3cm
\includegraphics*[width=0.4\textwidth]{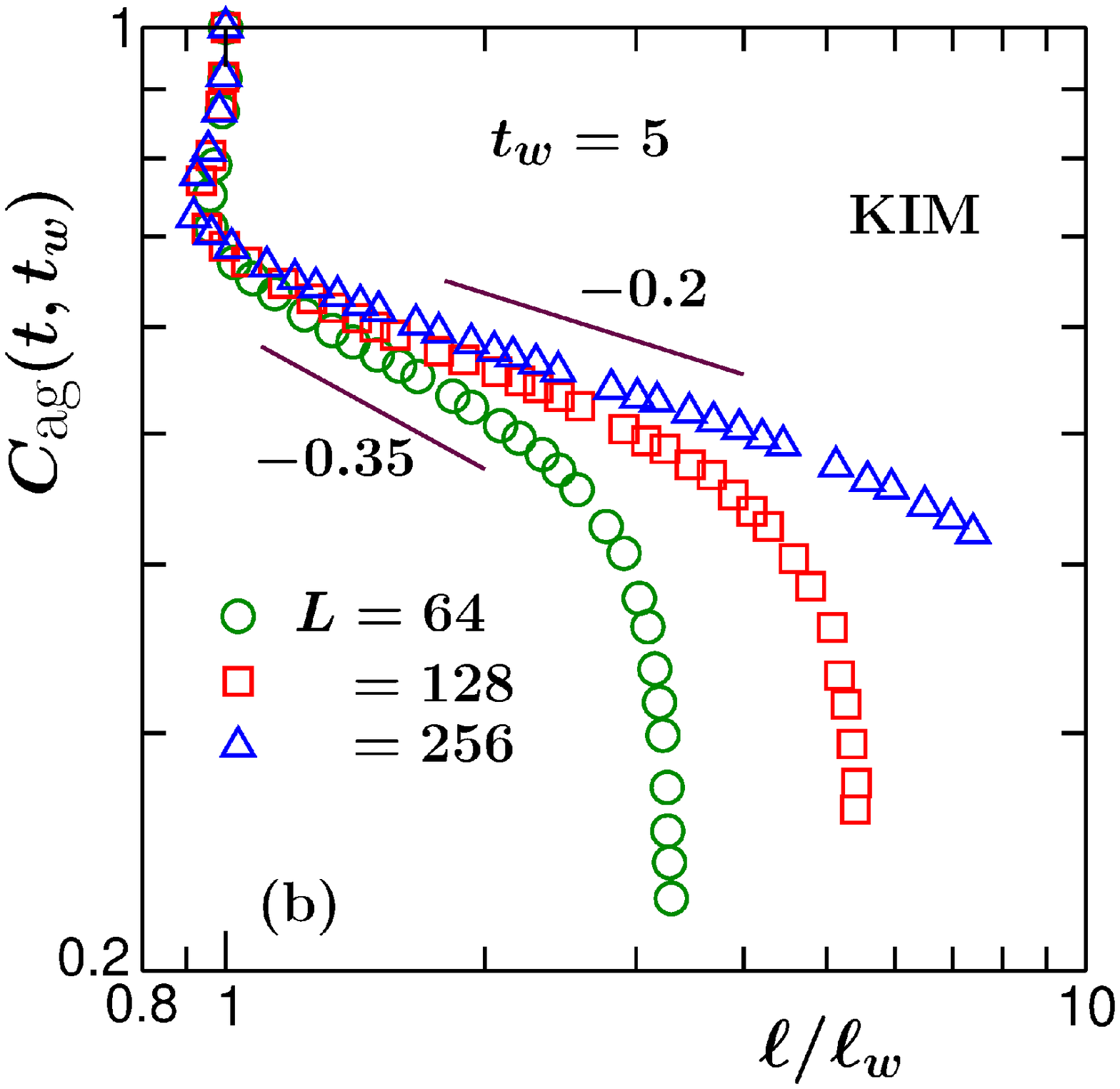}
\caption{\label{fig3} 
(a) Log-log plots of autocorrelation function, $C_{\textrm{ag}} (t,t_w)$, for the KIM, versus $\ell/\ell_w$. Data for a few different $t_w$ are shown. All results are for $L=256$. 
(b) Same as (a) but here we have fixed $t_w$ to $5$ and presented results for a few values of $L$. Inside both the frames the solid lines represent power laws. The values of the exponents are 
mentioned next to the lines.}
\end{figure}
\par
Following the discussion and presentation of results relevant for the scaling analysis of the aging data for the critical starting point, we now focus on the primary objective. 
In Fig. \ref{fig3} we present results for $C_{\textrm{ag}}(t,t_w)$, versus $\ell/\ell_w$, for the KIM. In part (a) we fix the system size and include data from few different $t_w$ values. 
On the other hand, in part (b) $t_w$ is fixed and $L$ is varied. In none of the cases collapse of data is observed. This should be contrasted with the available literature \cite{midya_jpcm,midya_pre} for quenches from $T_s = \infty$. Such non-scaling behavior for quenches from the critical point is because of the fact that for $L<\infty$, the structure, during evolution, quickly starts deviating from the desired scaling, as already mentioned. 
To overcome this problem we will perform extrapolation exercise to obtain the value of $\lambda$ in the $L=\infty$ limit. Note that very early-time structural change brings non-monotonicity in the length. This is reflected in the plots of Fig. \ref{fig3} for smaller values of $t_w$. During this period, we believe, the system is trying to arrive at the scaling structure. However, with the increase of time departure from this structure occurs, earlier for smaller systems. 

In both Fig. \ref{fig3} (a) and \ref{fig3} (b), a common feature is the following. Each of the data sets tend to stabilize to a power-law decay over a certain range of $\ell/\ell_w$, but deviates from 
it when $\ell$ approaches $L$, i.e., $\xi$. These stabilized exponent values are, however, different from each other in part (a) as well as in part (b). In part (a), this is because of the fact that the structure for each $t_w$ is different. Recall, we have already mentioned above that this is a nonequilibrium feature related to finite system of any particular size. On the other
hand, even though in the case of part (b) $t_w$ is fixed, here one has different finite-size effects for different $L$ to start with, owing to different initial $\xi$ for different $L$. 
\begin{figure}
\centering
\includegraphics*[width=0.4\textwidth]{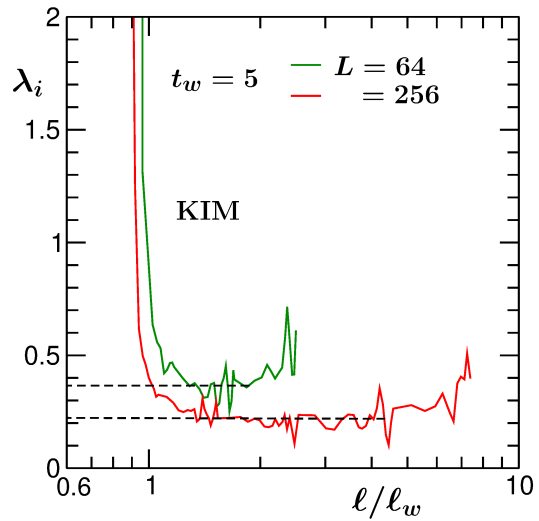}
\caption{\label{fig4} 
Instantaneous exponent $\lambda_i$ is plotted versus $\ell/\ell_w$, for the KIM, for two values of $L$. In each of the cases we have $t_w =5$. We extract $L$-dependent value, $\lambda_L$, from 
the flat regions of these plots.}
\end{figure}
\par
Nevertheless, for a fix $t_w$, with the increase of system size the exponents keep staying stable for longer ranges. Also, the rate of change of the exponent with the increase of $L$ keeps decreasing. That way one may like to consider a very large system to obtain $\lambda$ value that will be very close to that for $L=\infty$. We, however, would like to rely on an extrapolation method using relatively smaller systems. An advantage of using smaller systems is that one can get better statistics by running simulations with many independent initial configurations, using the same computational power that is needed to run single large system. Here note that reduction of error is not directly proportional to the size of a system \cite{sm_2010,heermann}.
\par
For the purpose of extrapolation, we need to obtain the exponent values in the stabilized regions accurately. For this we take help of the instantaneous exponent \cite{midya_pre,huse,amar}
\begin{equation}\label{lambda_i}
 \lambda_i = -\frac{d\ln C_{\textrm{ag}} (t,t_w)}{d\ln x};\,\, x=\frac{\ell}{\ell_w}.
\end{equation}
In Fig. \ref{fig4}, as illustration, we plot this quantity as a function of $x$, for the KIM, for two values of $L$, by fixing $t_w$ to $5$. The $L$ dependent exponent, $\lambda_L$, we obtain from the flat regions of the plots, that also correspond to the minima. One can justify this by taking a closer look at the behavior of $C_{\textrm{ag}} (t,t_w)$ in Fig. \ref{fig3}. 
We expect that $\lambda_L$ in the limit $L=\infty$ will have same convergence for all values of $t_w$, because of the following reasons. For the meaningful scaling evolution, in the $L=\infty$ limit the structure should obey certain self-similarity all along \cite{humayun,bray_humayun}. If so, the value of $\lambda$ should not be affected by the choice of $t_w$. Note that in such a situation the bound of Eq. (\ref{yrd_bound}) does not change. For finite $L$, of course, the situation is different, as discussed and being observed. However, the intended extrapolation is expected to lead us to the thermodynamic $\lambda$, same for all $t_w$. If this is the case and the corresponding $\lambda$ is different from that for $T_s=\infty$, like in the ferromagnetic case, it should give indirect evidence that there exist different structural scalings in the conserved case also for $T_s=\infty$ and $T_s=T_c$. 
\begin{figure}
\centering
\includegraphics*[width=0.4\textwidth]{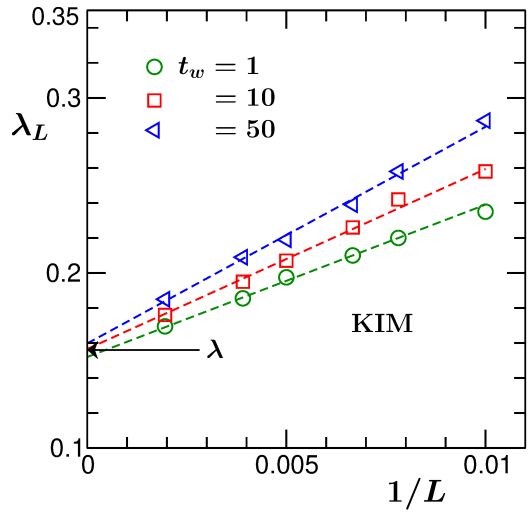}
\caption{\label{fig5} 
We have plotted $\lambda_L$ as a function of  $1/L$, for the KIM. Results for a few values of $t_w$ are included. The solid lines are linear fits for extracting $\lambda=\lambda_{L=\infty}$, value of which is marked by an arrow-headed line.}
\end{figure}
\par
Finally, to obtain the thermodynamic limit value, in Fig. \ref{fig5}  we have plotted $\lambda_L$, as a function of $1/L$, for a few values of $t_w$, again from the KIM. These multiple plots provide a good sense of convergence. From this exercise we quote
\begin{equation}\label{lambda_kim_tcl}
 \lambda = \lambda_{L=\infty} = 0.155 \pm 0.025.
\end{equation}
Since all the data sets appear linear, we have obtained the above quoted number from linear fittings. This number we compare with \cite{midya_pre}  $\lambda$ for KIM when $T_s = \infty$ in $d=2$, viz., 
\begin{equation}\label{lambda_kim}
 \lambda \simeq 3.6. 
\end{equation}
There exists huge difference between the quoted values in Eqs. (\ref{lambda_kim_tcl}) and (\ref{lambda_kim}).
\begin{figure}
\centering
\includegraphics*[width=0.4\textwidth]{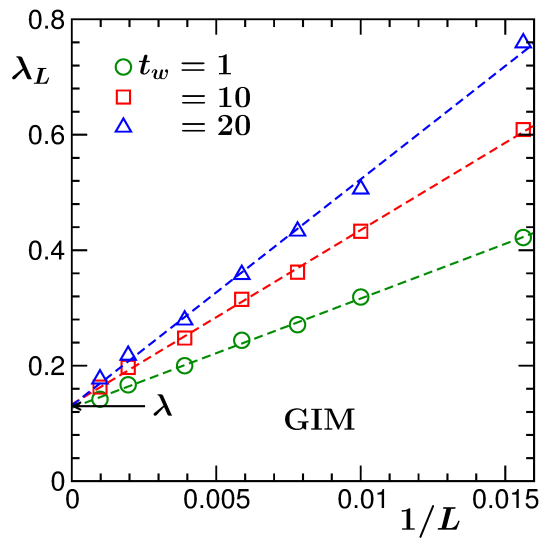}
\caption{\label{fig6} 
Same as Fig. \ref{fig5} but here we have presented results from the GIM.}
\end{figure}
\begin{figure}
\centering
\includegraphics*[width=0.4\textwidth]{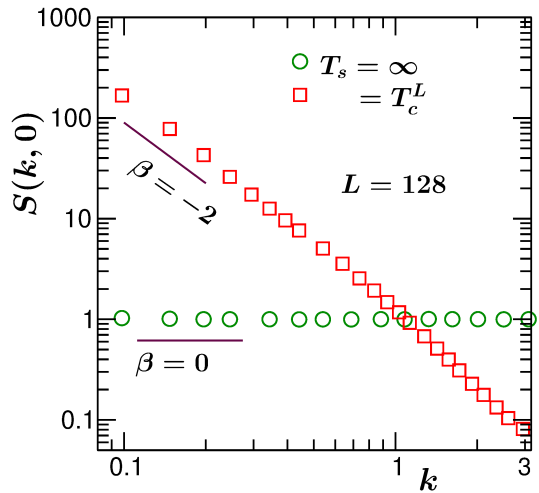}
\caption{\label{fig7} 
Plots of equal time structure factor $S(k,t_w=0)$ as a function of wave number $k$. We have shown results from $T_s = \infty$ and $T_s = T_c^L$. In each of the cases, we have used $L=128$. The solid lines represent the values of $\beta$.}
\end{figure}
\par
To validate our result of Eq. (\ref{lambda_kim_tcl}), we have applied the same method to the simulation data for the GIM. For this case, plots of $\lambda_L$ versus $1/L$, for different $t_w$ values, are shown in Fig. \ref{fig6}. Here also one can appreciate nice convergence of the data sets for different values of $t_w$. The corresponding value is
\begin{equation}\label{lambda_gim_tcl}
 \lambda = 0.13 \pm 0.01.
\end{equation}
This is certainly in extremely good agreement with the theoretical prediction \cite{humayun,bray_humayun}, viz., $\lambda = 0.125$. We mention here that in the previous simulation studies \cite{corberi_villa,humayun,bray_humayun} no such attempts have been made to estimate $\lambda$ for $T_s=T_c$ even for the GIM. Only checks for the consistency with the analytical theory were performed.
The outcome of this exercise here certainly puts confidence in the number quoted in Eq. (\ref{lambda_kim_tcl}). The number in Eq. (\ref{lambda_gim_tcl}), for GIM in $d=2$, should be compared with the corresponding value for $T_s=\infty$, which is $\simeq 1.3$. Thus, for both KIM and GIM, the values of $\lambda$ for $T_s=\infty$ and $T_s=T_c$ universality classes are vastly different. 

Next we aim at checking whether these numbers satisfy the YRD bound. For that purpose, in Fig. \ref{fig7} we have plotted  $S(k,0)$ as a function of $k$, on a log-log scale. We have included data from both 
$T_s=\infty$ and $T_s = T_c^L$, with $L=128$. In the case of $T_s=\infty$, flat behavior for the whole range of $k$ is observed. So, we have $\beta=0$.  
Naturally, $\lambda \simeq 3.6$ (for KIM) and $\lambda \simeq 1.3$ (for GIM) satisfy the corresponding bound, 
which is $\lambda\geq1$. 
However, for the conserved case, when scaling (overlap of data from different $t_w$) is observed starting from large $t_w$, for $T_s=\infty$, value of $\beta$ by then changes \cite{yeung,ahmed} to approximately $4$. In that case the bound becomes $\lambda \ge 3$. So, the estimate $\lambda \simeq 3.6$ still satisfies the modified bound in the scaling regime of $t_w$. In the nonconserved case, however, as already stated, $\beta$ remains zero for the whole evolution. 

It appears that the bounds are satisfied for $T_s = T_c^L$ also. In this case $\beta$ assumes a negative value, viz., $\beta\simeq -2$. Thus, the corresponding lower bound is below both the above quoted values, i.e., $\lambda\simeq0.16$ 
(for KIM) and $\lambda\simeq0.13$ (for GIM). We have verified that no violation occurs even with the progress of time.
\section{Conclusion}
We have presented results for aging phenomena in the two-dimensional Ising model \cite{fisher}. The results were obtained from Mote Carlo simulations \cite{binder_heermann,landau,dan_frenkel} with implementation of two different mechanisms. Our primary focus was on kinetics of phase 
separation in solid binary mixtures. For this we have used the Kawasaki exchange kinetics \cite{kawasaki}. For the purpose of verification of the adopted scaling method and thus, the outcome for the binary mixture, we have presented results for ordering in uniaxial ferromagnets as well, for which there exists theoretical prediction for comparison. 
In this case the results were obtained via the implementation of Glauber kinetics \cite{glauber}. Our objective was to estimate the aging exponent $\lambda$, related to the power-law decay of the order-parameter autocorrelation 
function \cite{fisher_huse} $C_{\textrm{ag}} (t,t_w)$, corresponding to the universality class \cite{humayun,bray_humayun} decided by quenches from $T_s=T_c$, for which one has infinitely correlated configurations \cite{fisher}.

For quenches from the critical point, simulation results suffer significantly from finite-size effects. This problem was appropriately taken care of by implementing finite-size scaling technique of equilibrium critical phenomena and devising an extrapolation method for analysis of the out-of-equilibrium data. We believe that our results are quite accurate for thermodynamically large systems.
\par
It appears that for both types of systems, viz., phase separating binary mixtures and ordering ferromagnets, the values of $\lambda$ for $T_s=T_c$ are drastically smaller than those for the universality class corresponding \cite{liu,midya_jpcm,midya_pre} to $T_s=\infty$. Nevertheless the obtained values for $T_s=T_c$ satisfy the lower bounds predicted by Yeung, Rao and Desai \cite{yrd}. To the best of our knowledge, these are the first results for solid mixtures, as far as quenches from $T_c$ is concerned. 
\par
In the case of ferromagnets already it was shown that the growth exponent remains same for the two above mentioned universality classes \cite{humayun,bray_humayun}. Our recent work \cite{nv_2} on growth for the KIM also points towards the same possibility. 
Overall, thus, it appears that there exists strong qualitative similarity between cases with conserved and non-conserved dynamics, as far as the universalities with respect to quenches from correlated and 
decorrelated initial configurations are concerned.
\par
Other important exponent that can be calculated for the binary mixture with both $T_s=\infty$ and $T_s=T_c$ is the one related to the decay of persistent probability \cite{bray_majumdar}. For this exponent, however, due to certain technical reasons \cite{derrida} quenches to very low temperature becomes necessary. In that case, for conserved dynamics, there exists severe problem with metastability. This makes the problem rather challenging.

\end{document}